\documentstyle[a4,epsf]{espart}

\begin{document}

\begin{frontmatter}

\title{Study of odd-mass $N=82$ isotones with realistic effective 
       interactions}

\author[jyv]{J.\ Suhonen},
\author[jyv]{J.\ Toivanen}, 
\author[oslo]{A.\ Holt}, 
\author[oslo]{T.\ Engeland}, 
\author[oslo]{E.\ Osnes} and
\author[kbh]{M.\ Hjorth-Jensen}
\address[jyv]{Department of Physics, University of Jyv\"{a}skyl\"{a},
              P.O.Box 35, FIN-40351 Jyv\"{a}skyl\"{a}, Finland}
\address[oslo]{Department of Physics, University of Oslo, N-0316 Oslo, Norway}
\address[kbh]{Nordita, Blegdamsvej 17, DK-2100 K\o benhavn \O, Denmark}

\maketitle

\begin{abstract}

The microscopic quasiparticle-phonon model, MQPM, is used to study
the energy spectra of the odd $Z=53 - 63$, $N=82$ isotones. 
The results are compared with experimental data, with the extreme
quasiparticle-phonon limit and with the results of an unrestricted 
$2s1d0g_{7/2}0h_{11/2}$ shell model (SM) calculation.
The interaction used in these calculations is a realistic 
two-body $G$-matrix interaction derived from modern meson-exchange 
potential models for the nucleon-nucleon interaction. 
For the shell model all the 
two-body matrix elements are renormalized by the $\hat{Q}$-box method 
whereas for the MQPM the effective interaction is defined by the $G$-matrix.

\end{abstract}

\end{frontmatter}



\section{Introduction}\label{sec:sec1}

The quasiparticle-phonon coupling (QPM) scheme is a convenient way of
exploiting the rich data on low-energy excitations of doubly-even nuclei
in studying nuclear-structure effects in odd-mass (odd-$A$ for short)
nuclei. Since it was introduced \cite{KIS63} it has been used extensively,
in various forms \cite{HAL67,DRE71,GUN74,HEL81,MAN90,DIA94,TOI95}, to discuss 
energy spectra of odd-$A$ nuclei. This scheme makes use of the property 
of the BCS quasiparticles being the elementary excitations of an odd-$A$ 
nucleus and assumes that by coupling them to the few lowest-energy (collective)
excitations of the doubly-even reference nucleus one is able to
describe, at least qualitatively, the spectroscopy of odd-$A$ nuclei.
In many cases only the first quadrupole and/or octupole phonon have been
used in the calculations.

In the traditional quasiparticle-phonon models (QPM) independent model 
hamiltonians have been used to create the quasiparticles, the phonons 
and their coupling. 
This means that the 
two-body interaction matrix elements of the many-fermion hamiltonian
are not internally consistent.
For the phonons one usually adopts very simple
model hamiltonians, like pairing-plus-quadrupole or semiempirical ones while
the quasiparticle energies can be extracted from simple pairing 
hamiltonians or directly from experiments. The quasiparticle-phonon 
coupling term has usually the simplest possible phenomenological form and may
have adjustable parameters to control the energy of the
quasiparticle-phonon multiplets.
In Ref.\ \cite{TOI95} a model was introduced, the so-called microscopic
quasiparticle-phonon model (MQPM), which bears resemblance to the
traditional QPM, but which uses a microscopic hamiltonian 
and a scheme to optimize the size of the
quasiparticle-phonon basis used in the calculations. In this sense, see the
discussion in section \ref{sec:sec2}, it represents also an improvement
over the traditional QPM since the quasiparticle-phonon interaction is 
treated in a more consistent way.

In order to test the MQPM method, we have singled out the 
$N=82$ isotones. 
The sequence of semi-magic $N=82$ nuclei 
shows a high degree of regularity which makes them well suited for 
systematic studies and for testing of microscopic nuclear models. 
Pairing effects seem to play an important role in the even nuclei, and the 
low-lying states in the odd nuclei may be described as 
one-quasiparticle states with increasing fermi level. Thus, models like the
MQPM may be viewed as a reasonable starting point for the description
of such nuclei.

The first aim of this work is therefore to compare 
the MQPM with the more traditional QPM and QRPA approaches in order to
see how well the experimental spectra can be reproduced and to interpret
eventual differences.  

A comprehensive study of the even $N=82$ isotones was carried out by us in a 
previous work \cite{HOL97}. There a comparison was made between 
the QRPA  and the results obtained with an extensive  shell model 
calculation. Energy spectra 
were generally well described and so were the 
transition probabilities. 
In Ref.\ \cite{HOL97} one of the aims was
to calculate an effective interaction based on
modern meson-exchange models for the nucleon-nucleon (NN) potential. 
The first step in the derivation of an effective interaction 
V$_{\rm eff}$ was to renormalize
the NN potential through the so-called $G$-matrix. The $G$-matrix was in turn
used in a perturbative many-body scheme, see e.g.
Ref.\ \cite{HJO95} for further details, to derive
an effective interaction appropriate for the $N=82$ isotones.
The effective interaction
is meant to take into account degrees of freedom not included in the model
space.
This  interaction 
was, in turn, applied in a full shell model  calculation 
with a model space consisting 
of the orbitals $2s_{1/2}$, $1d_{5/2}$, $1d_{3/2}$, $0g_{7/2}$ and $0h_{11/2}$ 
for the $Z=52 - 64$, $N=82$ isotones. 

The second aim of this work is therefore to extend the comparative 
analysis of Ref.\ \cite{HOL97}
for the even isotones to the case of the odd isotones. 
We will employ the
same effective interaction and model space in the shell model analysis. 
Moreover,
the same $G$-matrix used in the QRPA studies of Ref.\ \cite{HOL97} 
will be used in the MQPM calculations.  
Since the MQPM 
discussed here (and the QRPA method of \cite{HOL97} as well)   
employs a larger single-particle 
space than the perturbative many-body scheme, 
our hope is to see whether  the two approaches 
could shed light on different many-body contributions and 
their influence on various 
spectroscopic observables.

The $N=82$ isotones have previously been studied extensively by Heyde
and Waroquier \cite{WAR70,HEY71} by using the quasiparticle Tamm-Dancoff 
approximation including one- and three-quasiparticle states as basis
states and projecting out the spurious three-quasiparticle components.
The calculations were done using the surface-delta interaction
\cite{WAR70} and a more elaborate phenomenological interaction
\cite{HEY71} in the proton $2s_{1/2}1d_{5/2}1d_{3/2}0g_{7/2}0h_{11/2}$ 
single-particle basis, i.e. in the same basis which we adopt for the
shell model in our present article. However, the more realistic
results of Ref. \cite{HEY71} are not accessible for direct comparison with
the present results of the MQPM for the following reasons: A) the MQPM
employs a larger valence space where also neutron degrees of freedom
are included. B) In the MQPM we use the same proton single-particle
energies, extracted from the experimental spectrum of $^{133}$Sb, for
all the odd isotones whereas in Ref. \cite{HEY71} the inverse-gap-equation
method was used to extract single-particle energies from the 
experimental spectra of each odd $N=82$ isotone separately. C) In
Ref. \cite{HEY71} the phenomenological force, namely a central force of
Gaussian shape with spin exchange, was fitted from case to case by
the above-mentioned inverse-gap-equation method, and overall by level
schemes and transition rates of the even $N=82$ isotones. 
Contrary to this, in the present MQPM calculation the same bare 
G-matrix interaction is used for all the even and odd $N=82$ isotones
without any fitting procedures. For the above reasons we refrain from
direct comparison with the results of Ref. \cite{HEY71} in this article
and, instead, concentrate on comparison with the shell model results.
 
The theoretical framework of
the MQPM is presented in section \ref{sec:sec2}. 
A brief review of the effective
interaction theory and the shell model follows in 
section \ref{sec:sec3}. The results are presented and discussed in section 
\ref{sec:sec4}. Finally, in section \ref{sec:sec5} we draw our conclusions.

\section{Theoretical framework of the MQPM}\label{sec:sec2}

The microscopic quasiparticle-phonon model, MQPM, represents a 
scheme to treat all the three parts
of the hamiltonian, namely the quasiparticle, phonon and 
quasiparticle-phonon terms, on equal footing. This is possible by 
starting from a microscopic hamiltonian with two-body matrix elements 
derived from effective matrix elements such as a 
$G$-matrix. The $G$-matrix is a medium modified nucleon-nucleon
interaction where all ladder type diagrams are summed to infinite
order.
This method enables one in a systematic way to derive both 
the proton-neutron and the like-nucleon two-body interaction. 
In the 
quasiparticle language these parts of the interaction relate to the
$H_{31}$ and $H_{22}$ parts of the quasiparticle representation
of the nuclear hamiltonian. 
The $H_{22}$ part is treated in the BCS and
quasiparticle random-phase approximation (QRPA) framework and leads
to definition of the quasiparticles and the excitation (phonon) 
spectrum of the doubly-even reference nucleus. The $H_{31}$ part is then 
diagonalized in the quasiparticle-phonon basis discussed below.

The MQPM treats the structure
of the odd-$A$ nuclei in four steps. First, the neighboring even-even
nucleus, or nuclei, can be used to study the properties of the chosen
mass region and to fix the possible free parameters of the model
hamiltonian. In the present case, as also in \cite{HOL97},
we have used a $G$-matrix derived from modern meson-exchange
potential models, 
and thus no phenomenological renormalization of the two-body interacton 
was done. This hamiltonian is used to generate the phonons which
are excitations of the even-even nuclei. In the MQPM the phonons are
derived by the use of the quasiparticle
random-phase approximation (QRPA) procedure \cite{RIN80}. Second, the 
monopole part of the same hamiltonian is used to generate the 
quasiparticles, which are the 
basic building blocks of the odd-$A$ excitations,
through the BCS procedure. As the third step, the two basic 
excitations, QRPA phonons and BCS quasiparticles, are coupled to 
form a wave function basis for a realistic treatment of the 
odd-$A$ nucleus. As the last step, the residual hamiltonian, 
containing the interaction of the odd nucleon with the even-even 
reference nucleus (the $H_{31}+H_{13}$ part of the hamiltonian in Eq.\ (3)
below) is diagonalized in this (over-complete) basis.

In the MQPM the starting point is the $A$-fermion hamiltonian
\begin{equation}
       H = \sum_{\alpha} \varepsilon_{\alpha} c^{\dagger}_{\alpha} c_{\alpha} +
       \frac{1}{4}\sum_{\alpha\beta\gamma\delta} 
       {\bar{v}}_{\alpha\beta\gamma\delta}
       c^{\dagger}_{\alpha}c^{\dagger}_{\beta} c_{\delta} c_{\gamma},
\end{equation}
containing antisymmetrized two-body matrix elements 
${\bar{v}}_{\alpha\beta\gamma\delta}=\langle \alpha\beta\vert v
\vert \gamma\delta\rangle - \langle \alpha\beta\vert v 
\vert \delta\gamma\rangle$
obtained from the Bonn-A $G$-matrix. Greek 
indices denote all single-particle quantum numbers 
$\alpha=\lbrace a,m_a \rbrace$, and roman indices, when used, 
denote all single-particle quantum numbers except the magnetic ones, 
i.e. $a=\lbrace n_a,l_a,j_a \rbrace$. 

The approximate ground state of the even-even reference nucleus
is obtained from a BCS calculation, where
quasiparticle energies and occupation factors $u_a$ and $v_a$ are
obtained from the Bogoliubov-Valatin transformation to quasiparticles

\begin{eqnarray}
       a^{\dagger}_{\mu} &=& u_{\mu}c^{\dagger}_{\mu}-v_{\mu}\tilde{c}_{\mu} 
       \nonumber \\
       \tilde{a}^{\dagger}_{\mu} &=& u_{\mu}\tilde{c}^{\dagger}_{\mu}
       +v_{\mu}c_{\mu}\,, 
\end{eqnarray}
where $\tilde{a}^{\dagger}_{\mu}=a^{\dagger}_{-\mu}(-1)^{j+m}$ and 
$\tilde{c}^{\dagger}_{\mu}=c^{\dagger}_{-\mu}(-1)^{j+m}$.
After this transformation the hamiltonian can be written in the
form
\begin{equation}
       H = \sum_{\alpha} E_a a^{\dagger}_{\alpha} a_{\alpha} + H_{22} + H_{40}
           + H_{04} + H_{31} + H_{13},
\end{equation}
where $E_a$ are the quasiparticle energies and other terms
of the hamiltonian are normal-ordered parts of the residual
interaction labeled according to the number of quasiparticle
creation and annihilation operators which
they contain \cite{SUH88}.

In the first version of the MQPM \cite{TOI95} the coupling part of
the microscopic hamiltonian, $H_{31}$,  does not emerge from
the equations-of-motion method (EOM) \cite{ROW68}. The EOM method
introduces an additional term into the quasiparticle-phonon
matrix elements, not taken into account in Ref. \cite{TOI95}.
In the present article we use the EOM form of the coupling
part of the hamiltonian. In addition of being microscopically more
justified, this form of the coupling hamiltonian yields results
closer to the experimental data and thus improves the quantitative 
predictibility of the MQPM. 

In the MQPM calculation we use for the protons the 1p0f and 2s1d0g 
oscillator major shells supplemented by the h$_{11/2}$ intruder
orbital from the next oscillator major shell. For the neutrons we
use the 2s1d0g$_{7/2}$ and 2p1f0h valence space.
The proton single-particle energies $\varepsilon_{\alpha}$ of Eq.\
(1)
are taken to correspond to the ones of Fig.\ 1, 
i.e.\ we take the same 
relative spacing of the key orbitals as used in the shell model
calculation. The single-particle energies of Fig.\ 1 are extracted 
from the experimental $^{133}$Sb spectrum \cite{STO95}, 
with exception of the $2s_{1/2}$ single-particle energy 
which has not yet been measured. We have used the same value for the 
$2s_{1/2}$ single-particle energy as in the 
work of Sagawa et al. \cite{SAG87}.
The proton energies outside this set of states, as well as
the neutron single-particle energies, we take from 
the Coulomb-corrected Woods-Saxon potential with the
parametrization of \cite{BOH69}. These single-particle energies
are displayed in Tables \ref{tab:table1} and \ref{tab:table2}. 
\begin{table}
\caption{Proton single-particle energies used in 
the MQPM calculations. The single-particle 
energies of the rest of the proton valence space 
are given in Fig.\ 1.}

\begin{center}
\begin{tabular}{ccccccc}
\hline
s.p. orbital & $^{135}$I & $^{137}$Cs & $^{139}$La & 
$^{141}$Pr & $^{143}$Pm & $^{145}$Eu \cr
\hline 
0f$_{7/2}$ & -12.289 & -12.114 & -11.945 & -11.782 & -11.619 & -11.469 \\
0f$_{5/2}$ & -8.808 & -8.694 & -8.583 & -8.476 & -8.367 & -8.269 \\
1p$_{3/2}$ & -7.597 & -7.443 & -7.294 & -7.150 & -7.005 & -6.872 \\
1p$_{1/2}$ & -6.212 & -6.083 & -5.959 & -5.838 & -5.717 & -5.618 \\
0g$_{9/2}$ & -5.402 & -5.308 & -5.217 & -5.131 & -5.043 & -4.967 \\
0g$_{7/2}$ & 0.000 & 0.000 & 0.000 & 0.000 & 0.000 & 0.000 \\
\hline
    
\end{tabular}

\end{center}
\label{tab:table1}\end{table}

\begin{table}
\caption{
Neutron single-particle energies used in the MQPM calculations.}
\begin{center}
\begin{tabular}{ccccccc}
\hline
s.p. orbital & $^{135}$I & $^{137}$Cs & $^{139}$La & 
$^{141}$Pr & $^{143}$Pm & $^{145}$Eu \cr
\hline 
1d$_{5/2}$ & -7.192 & -7.251 & -7.302 & -7.345 & -7.382 & -7.413 \\
0g$_{7/2}$ & -7.156 & -7.283 & -7.399 & -7.504 & -7.600 & -7.688 \\
2s$_{1/2}$ & -5.402 & -5.425 & -5.443 & -5.456 & -5.466 & -5.473 \\
1d$_{3/2}$ & -5.151 & -5.203 & -5.249 & -5.289 & -5.325 & -5.356 \\
0h$_{11/2}$ & -4.226 & -4.341 & -4.444 & -4.537 & -4.621 & -4.696 \\
1f$_{7/2}$ & 0.000 & 0.000 & 0.000 & 0.000 & 0.000 & 0.000 \\
2p$_{3/2}$ & 1.358 & 1.499 & 1.624 & 1.735 & 1.834 & 1.923 \\
0h$_{9/2}$ & 1.724 & 1.602 & 1.488 & 1.379 & 1.275 & 1.176 \\
2p$_{1/2}$ & 2.166 & 2.371 & 2.549 & 2.704 & 2.841 & 2.963 \\
1f$_{5/2}$ & 2.715 & 2.792 & 2.853 & 2.903 & 2.945 & 2.979 \\
\hline
    
\end{tabular}

\end{center}
\label{tab:table2}
\end{table}

\begin{figure}[htbp]
      \setlength{\unitlength}{1.4cm}
      \begin{center}
      \begin{picture}(2,5)(0,-1)
      \newcommand{\lc}[1]{\put(0,#1){\line(1,0){1}}}
      \newcommand{\ls}[2]{\put(2,#1){\makebox(0,0){{\scriptsize $#2$}}}}
      \newcommand{\lsr}[2]{\put(2,#1){\makebox(0,0){{\scriptsize $#2$}}}}
      \put(-.25,3.4){\makebox(0,0){\large MeV}}
      \thicklines
      \put(-.75,-.5){\line(0,1){4}}
      \multiput(-.75,.0)(0,1){4}{\line(1,0){.1}}
      \multiput(-.75,.5)(0,1){3}{\line(1,0){.05}}
      \put(-1.,3){\makebox(0,0){3}}
      \put(-1.,2){\makebox(0,0){2}}
      \put(-1.,1){\makebox(0,0){1}}
      \put(-1.,0){\makebox(0,0){0}}
      \lc{0.000}   \ls{0.000000}{7/2+ \;\;0.000}
      \lc{0.962}   \ls{0.962}{5/2+ \;\;0.962}
      \lc{2.708}   \ls{2.608}{3/2+ \;\;2.708}
      \lc{2.792}   \ls{2.792}{11/2- \;\;2.792}
      \lc{2.990}   \ls{2.990}{1/2+ \;\;2.990}
      \end{picture}
      \end{center}
      \caption{Adopted single-particle energies for the proton orbitals 
        $2s_{1/2}$, $1d_{3/2}$, $1d_{5/2}$, $0g_{7/2}$ and $h_{11/2}$ in the 
        MQPM and the shell model calculation.}
      \label{fig:sp-energies}
\end{figure}
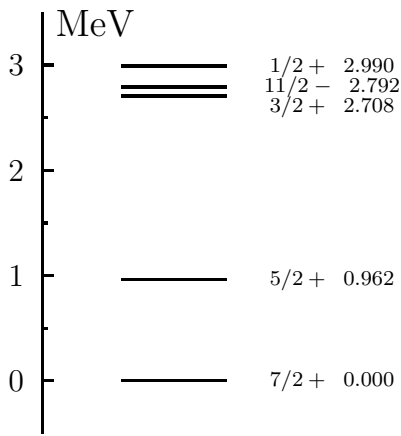

In the next step a correlated ground state and the
excited states of the even-even reference nucleus are constructed
by use of the QRPA. In the QRPA the creation operator for an excited
state (QRPA phonon) has the form

\begin{equation}
      Q^{\dagger}_{\omega} = \sum_{a\le a'} \left\lbrack X^{\omega}_{aa'}
      A^{\dagger}(aa';J_{\omega}M) -
      Y^{\omega}_{aa'} {\tilde{A}}(aa';J_{\omega}M) \right\rbrack,
\end{equation}

where the quasiparticle pair creation and annihilation operators
are defined as
$A^{\dagger}(aa';JM)=\sigma^{-1}_{aa'}\Bigl\lbrack a^{\dagger}_{a}
a^{\dagger}_{a'} 
\Bigr\rbrack_{JM}$,  
${\tilde{A}}(aa';JM)=\sigma^{-1}_{aa'}\Bigl\lbrack \tilde{a}_{a}\tilde{a}_{a'} 
\Bigr\rbrack_{JM}$ and $\sigma_{aa'}=\sqrt{1+\delta_{aa'}}$.
Here the greek indices $\omega$ denote phonon spin $J$ and
parity $\pi$. Furthermore, they contain
an additional quantum number $k$ enumerating the different QRPA
roots for the same angular momentum and parity. Thus
$\omega=\lbrace J_{\omega},\pi_{\omega},k_{\omega} \rbrace$.

For each value of the angular momentum and parity the spectrum
of the even-even nucleus is constructed by diagonalizing the QRPA
matrix containing the usual submatrices $A$ (quasiparticle-quasiparticle
interaction) and $B$ (induced by correlations of the ground state)
\cite{RIN80}. The basis states in our quasiparticle-phonon calculation are
constructed from the previously determined BCS quasiparticles and
the QRPA phonons. In the MQPM we make the following ansatz for the 
states in an even-odd nucleus

\begin{equation}
      \vert i;jm\rangle = 
      \left(\sum_n C^i_n a^{\dagger}_{njm} + \sum_{nj'\alpha}
      C^i_{nj'\alpha} \Bigl\lbrack a^{\dagger}_{nj'} Q^{\dagger}_\alpha 
      \Bigr\rbrack_{jm} \right)
      \vert -\rangle,
\end{equation}

\noindent
where $\vert -\rangle$ denotes the QRPA vacuum of the even-even reference
nucleus. The overlap matrix elements between
the quasiparticle-phonon states and the matrix elements of the
quasiparticle-phonon hamiltonian in this basis have the following form
\begin{eqnarray}
      \langle -\vert \Bigl\lbrack a^{\dagger}_{n} 
      Q^{\dagger}_{\alpha} \Bigr\rbrack^{\dagger}_{j}
      \Bigl\lbrack a^{\dagger}_{n'} Q^{\dagger}_{\alpha'} 
      \Bigr\rbrack_{j} \vert -\rangle 
      &=& \delta_{\alpha\alpha'} \delta_{nn'} + 
      A\left(\alpha n\alpha' n';j\right), \nonumber \\
      \langle -\vert \Bigl\lbrack a^{\dagger}_{n} Q^{\dagger}_{\alpha}
      \Bigr\rbrack^{\dagger}_{j} H \Bigl\lbrack a^{\dagger}_{n'} 
      Q^{\dagger}_{\alpha'}
      \Bigr\rbrack_{j} \vert -\rangle 
      &=& \frac{1}{2}\left( \hbar\Omega_{\alpha} + 
      E_n + \hbar\Omega_{\alpha'} + E_{n'}\right) 
      \langle -\vert \Bigl\lbrack a^{\dagger}_{n} Q^{\dagger}_{\alpha} 
      \Bigr\rbrack^{\dagger}_{j}
      \Bigl\lbrack a^{\dagger}_{n'} Q^{\dagger}_{\alpha'} 
      \Bigr\rbrack_{j} \vert -\rangle \nonumber \\
      &-& \frac{1}{2} {\hat J_{\alpha}}{\hat J_{\alpha '}} \sum_a
      {\left\lbrace\matrix{j_{n'}&j_a&J_\alpha\cr j_n&j&J_{\alpha'} 
      \cr}\right\rbrace} \\
      &\times& \left( \hbar\Omega_{\alpha} + E_{n} +
       \hbar\Omega_{\alpha'} + E_{n'} - 2E_{a} \right) {\bar X}^{\alpha}_{an'} 
      {\bar X}^{\alpha'}_{an} \sigma^{-1}_{an}\sigma^{-1}_{an'}, \nonumber
\end{eqnarray}
where $\Omega_{\alpha}$ denote the QRPA-phonon energies, and
\begin{equation}
      A\left(\alpha n\alpha' n';j\right) = {\hat J_{\alpha}}
      {\hat J_{\alpha '}} \sum_a
       \left\lbrack 
       {\left\lbrace\matrix{j_{n'}&j_a&J_\alpha\cr j_n&j&J_{\alpha'} 
      \cr}\right\rbrace}
      {\bar X}^{\alpha}_{an'} {\bar
      X}^{\alpha'}_{an} - {\delta_{jj_a}\over{{\hat j}^2}} {\bar
      Y}^{\alpha}_{an} {\bar Y}^{\alpha'}_{an'} \right\rbrack \sigma^{-1}_{an}
      \sigma^{-1}_{an'}\,.
\end{equation}
Here ${\bar X}^{\alpha}_{aa'}
\equiv X^{\alpha}_{aa'} - (-1)^{j_a+j_{a'}-J_{\alpha}}X^{\alpha}_{a'a}$.
The same definition holds for $\bar Y$.

The interaction matrix elements between the
one-quasiparticle and quasiparticle-phonon states have the following form:
\begin{eqnarray}
       \langle -\vert \Bigl\lbrack Q_{\alpha} a_n \Bigr\rbrack_{jm}
       {\hat{H}} a^{\dagger}_{\nu'} \vert -\rangle  &=& 
       \frac{1}{3} {\hat{J}_{\alpha}\over \hat{j}_{n'}}
       \sum_{a\le a'} H_{\rm pp}(aa'nn'J_{\alpha}) \left( u_a u_{a'}
       X^{\alpha}_{aa'} -v_a v_{a'} Y^{\alpha}_{aa'} \right) 
       \sigma_{aa'}^{-1} \nonumber \\
       &-& \frac{1}{3}{\hat{J}_{\alpha}\over{\hat{j}_{n'}}} \sum_{a\le a'}
       H_{\rm hh}(aa'nn'J_{\alpha})
       \left( v_a v_{a'} X^{\alpha}_{aa'} - u_a u_{a'} Y^{\alpha}_{aa'}
       \right) \sigma_{aa'}^{-1} \nonumber \\
       &+& \frac{1}{3}{\hat{J}_{\alpha}\over{\hat{j}_{n'}}} \sum_{a\le a'} 
       H_{\rm ph}(aa'nn'J_{\alpha}) \left( u_a v_{a'}
       X^{\alpha}_{aa'} + v_a u_{a'}
       Y^{\alpha}_{aa'} \right) \sigma_{aa'}^{-1} \\
       &-& \frac{1}{3}{\hat{J}_{\alpha}\over{\hat{j}_{n'}}}
       \sum_{a\le a'} H_{\rm hp}(aa'nn'J_{\alpha})
       \left( v_a u_{a'} X^{\alpha}_{aa'} + u_a v_{a'} Y^{\alpha}_{aa'}
       \right) \sigma_{aa'}^{-1}, \nonumber
\end{eqnarray}
where
\begin{eqnarray}
       H_{\rm pp} (nn'aa'J) &=& 2 v_n u_{n'} G(nn'aa'J)\,, \\
       H_{\rm hh}(nn'aa'J) &=& 2 u_n v_{n'} G(nn'aa'J)\,, \\
       H_{\rm ph}(nn'aa'J) &=& 2 v_n v_{n'} F(nn'aa'J) + 2 u_n u_{n'} 
       F(n'naa'J)(-1)^{j_n+j_{n'}+J}\,, \\
       H_{\rm hp}(nn'aa'J) &=& 2 u_n u_{n'} F(nn'aa'J) + 
       2 v_n v_{n'} F(n'naa'J)(-1)^{j_n+j_{n'}+J}.
\end{eqnarray}
In the previous version of our model \cite{TOI95} the second term in the
quasiparticle-phonon matrix elements of the hamiltonian in Eq. (6) was missing.
This additional term stems from the use of the equations-of-motion (EOM)
method \cite{ROW68} when deriving the eigenvalue equation (13) below.
It has an important effect on the location of the
three-quasiparticle-type states relative to the one-quasiparticle-type ones,
and it is essential for yielding theoretical results in agreement with data.

The overlap between the one-quasiparticle and the
quasiparticle-phonon states is always zero. However, the overlap
between two quasiparticle-phonon states can be non-zero and the
quasiparticle-phonon states form a non-orthogonal over-complete
basis set. The ansatz (5) leads to a generalized hermitian (or real and
symmetric) eigenvalue problem which has the form \cite{TOI95}
\begin{equation}
      \sum_j H_{ij} C^{\left(n\right)}_j = \lambda_n \sum_j S_{ij}
      C^{\left(n\right)}_j,
\end{equation}
where $H_{ij}=\langle i\vert H \vert j\rangle$ and 
$S_{ij}=\langle  i\vert j\rangle$
is the overlap matrix element between two basis states (one-quasiparticle
or quasiparticle-phonon states). To solve this rather involved
eigenvalue problem we adopt the method where we first solve the
eigenvalue equation for the overlap matrix ${\bf S}$:
\begin{equation}
      \sum_j S_{ij} u^{\left(k\right)}_j = n_k u^{\left(k\right)}_i.
\end{equation}
The eigenvectors can be written in the basis
$\lbrace \vert i\rangle\rbrace$ as
\begin{equation}
      \vert \tilde{k}\rangle=
      {1\over\sqrt{n_k}}\sum_i u^{\left(k\right)}_i \vert i\rangle.
\end{equation}
They have the property of being mutually orthogonal,
have a norm equal to unity and form a complete set after removing
states having eigenvalue $n_k=0$ (this removes the overcompleteness
of the set $\lbrace \vert i\rangle\rbrace$). 

Using the new orthogonal complete set of states (15) we can
transform (13) to an ordinary real and symmetric eigenvalue problem of
the form
\begin{equation}
       \sum_j \langle \tilde{i}\vert H \vert \tilde{j}\rangle 
        g^{\left(n\right)}_j = 
       \lambda_n g^{\left(n\right)}_i,
\end{equation}
where
\begin{equation}
      \langle \tilde{i}\vert H \vert \tilde{j}\rangle = 
      {1\over\sqrt{n_i n_j}} \sum_{kl}
      u^{\left(i\right)*}_k \langle k\vert H \vert l\rangle 
      u^{\left(j\right)}_l.
\end{equation}
The coefficients of the eigenstates are calculated from the $g$ coefficients
in the following way:
\begin{equation}
      C^n_i = \sum_k n^{-1/2}_k g^{\left(n\right)}_k
      u^{\left(k\right)}_i.
\end{equation}

In practice one omits states having eigenvalue $n_k$ less
than some set upper limit $\epsilon$.

In the present work we have constructed the odd-$A$ isotones by adding
a proton particle to the adjacent even $N=82$ isotone. 
The even $N=82$ isotones thus provide our quasiparticles and QRPA phonons 
to be used in the ansatz of Eq.\ (5). 
As a matter of fact, these phonons have
already been constructed in our previous study of the $N=82$ isotones
\cite{HOL97}. In the diagonalization of the MQPM
matrix of Eq.\ (13) we have used the 4-6 lowest QRPA phonons of
multipolarity $2^+$, $3^-$, $4^+$, $5^-$, $6^+$ and $7^-$. This number
of selected multipolarities is large enough to stabilize the spectrum
of the odd isotones (see Ref.\ \cite{TOI95}) and is referring to the
determination of the most relevant part of the
three-quasiparticle hilbert space mentioned in the introduction.

In this context it is important to point out that no $0^+$ phonons
are used to produce the MQPM results discussed in section
\ref{sec:sec4},
although in the first part of our work \cite{HOL97} the excited $0^+$ 
states were discussed extensively. There are two reasons why we omit
the $0^+$ phonons in the present MQPM calculation, namely, for the first, 
in the isotones $^{136}$Xe and $^{138}$Ba the calculated
energies of the first excited $0^+$ state, $0^+_2$, is seen to be more
than 1 MeV too low in comparison with the corresponding experimental
energy. In fact, in the QRPA the $0^+_2$ energy is lower than the 
$2^+_1$ energy for these isotones. Thus the wave function of the $0^+_2$
might not be adequately described
and this particular phonon would produce two
"spurious" low-energy excitations with spins $7/2^+$ and $5/2^+$ coming
from the $0^+_2\otimes 7/2^+$ and $0^+_2\otimes 5/2^+$ quasiparticle-phonon
couplings. Due to this unrealistic feature the $0^+$ phonons are discarded
in these isotones. Secondly, in the other isotopes under discussion
the calculated energy of the $0^+_2$ state is well above the energy
of the $2^+_1$ state (confirmed by the experimental data in the cases of
$^{140}$Ce, $^{142}$Nd and $^{144}$Sm) and calculations show that in
this case the inclusion of the $0^+$ phonons has only a negligible
effect on the results presented in section \ref{sec:sec4}.

\section{Effective interaction and the shell model}\label{sec:sec3}

Here we will briefly sketch the theory of the effective interaction and
the shell model (SM).

Our scheme to derive an effective interaction can be divided into three
steps. First one needs a free NN interaction V which is suitable
for nuclear physics at low and intermediate energies. At present, the
most viable approach seems to be the meson-exchange picture. Among the
meson-exchange models, one of the more successful ones is the 
one-boson-exchange model of the Bonn group \cite{MAC89}.

As the next step a reaction matrix $G$ is introduced. In this way we 
overcome the problem that the NN potential has a strongly 
repulsive core which makes 
it unsuitable for perturbative approaches. In this work we have calculated
the $G$-matrix using the so-called double-partitioning scheme, see 
Ref.\  \cite{HJO95} for a recent review.

The last step is to define a two-body interaction in terms of the $G$-matrix.
We include all diagrams to third order in perturbation theory and sum the
so-called folded diagrams to infinite order, an approach known as the folded 
diagram method, see e.g., Refs.\ \cite{HOL97,HJO95} for further details.

Our basic approach in solving the many-body eigenvalue problem is the 
Lanczos algorithm. This is an iterative method which was first applied to
nuclear physics problems by Whitehead \cite{WHI77}.
The eigenstates are expanded in an $m$-scheme slater determinant basis, 
which implies that the dimension of the problem grows
rapidly with increasing number of valence particles, see Table 
\ref{tab:dimension}. 
The advantage of
this representation is however the very efficient implementation of the 
computer code. This is a shell model calculation where no truncations 
of configurations are 
made. For more details of the shell model algorithm, see Ref.\ \cite{ENG95}.
\begin{table}[htbp]
\begin{center}
\caption{Number of basis states for the shell model calculation of the $N=82$
isotones, with the 
$1d_{5/2}$, $0g_{7/2}$, $1d_{3/2}$, $2s_{1/2}$ and $0h_{11/2}$
single particle orbitals.}
\begin{tabular}{lr|cr|lr}
\\\hline
System & Dimension & System & Dimension & System & Dimension \\
\hline
$^{134}$Te & 36        & $^{139}$La & 108 297   & $^{144}$Sm & 6 210 638 \\
$^{135}$I  & 245       & $^{140}$Ce & 323 682   & $^{145}$Eu & 9 397 335 \\ 
$^{136}$Xe & 1 504     & $^{141}$Pr & 828 422   & $^{146}$Gd & 12 655 280 \\
$^{137}$Cs & 7 451     & $^{142}$Nd & 1 853 256 & $^{147}$Tb & 15 064 787 \\
$^{138}$Ba & 31 124    & $^{143}$Pm & 3 609 550 & $^{148}$Dy & 16 010 204 \\
\hline
\end{tabular}
\label{tab:dimension}
\end{center}
\label{tab:table3}\end{table}

The model space 
for the shell model calculation and the effective interaction
is defined by the $N=4$ oscillator shell ($1d_{5/2}, 0g_{7/2},
1d_{3/2}, 2s_{1/2}$). In addition we have included the intruder $0h_{11/2}$ 
orbital from the $N=5$ oscillator shell. Our model space consists of the 
proton orbitals outside  the $^{132}$Sn core, ranging from the closed $Z=50$, 
$N=82$ core to the $Z=N=82$ core. There are no neutron degrees of freedom 
involved in this model. These degrees of freedom are accounted for by the 
various terms of the perturbation expansion used to
derive the effective interaction. 
The adopted single-particle spectrum is as displayed in Fig.\ 
\ref{fig:sp-energies}.

Before we present our comparison between the shell model results and
those obtained with the MQPM, we would like to draw the attention
to differences between the two methods.
 From the discussion in section \ref{sec:sec2} we recall that 
the MQPM method employs exactly the same $G$-matrix 
interaction as the one used to derive the shell model effective interaction.
The single-particle energies for protons in the orbitals  
$1d_{5/2}0g_{7/2}1d_{3/2}2s_{1/2}0h_{11/2}$ defining the shell model
space are the same as those used in the MQPM. In addition, no phenomenological
adjustments are made of the $G$-matrix in the MQPM approach. 
However, the reader should note that the single-particle basis
for protons is larger for the MQPM, allowing thereby for proton
core excitations across the $Z=50$ shell gap. Secondly, 
also neutrons are active,  yielding neutron core excitations across the 
$N=82$ shell gap. This might become important for the description of some 
low-energy collective excitations of the even $N=82$ isotones. In the shell 
model approach these degrees of freedom are supposed to be accounted for by 
terms included in the perturbative expansion of the effective interaction.
Substantial differences in the two approaches may therefore
reveal that such low-energy collective excitations are not accounted
for in the shell model approach.

\section{Results and discussion}\label{sec:sec4}

In Tables \ref{tab:odd-2}-\ref{tab:odd-7} we present our results 
for the calculated energy spectra
of the odd $N=82$ isotones of interest in this article. 
In these tables
we also present the available experimental data in the 
two first columns of the tables. It is to be noted that 
the spin assignments on the first
column are experimental ones and the spins in parentheses are only 
tentative. The MQPM and the SM results are always 
presented in the last
two columns of the tables.
The three middle colums are reserved for presenting 
the results concerning
the extreme quasiparticle-phonon picture. In this picture the energy
of the multiplet emerging from angular-momentum coupling 
of the $n$:th
phonon of spin $J$ and parity $\pi$, $J_n^{\pi}$, with the 
single-quasiparticle state $j^{\pi}$ is obtained by simply summing the 
phonon and the quasiparticle energies. The column "proposed config."
lists the phonon-quasiparticle states\footnote{In the extreme 
quasiparticle-phonon picture we assume that only the proton-quasiparticle 
states are active in the low-energy spectrum of the odd isotones.} which 
are among the leading ones in the wave function of the MQPM. 

In the first subsection below we discuss the differences between the
extreme quasiparticle picture and the MPQM. Thereafter, the MQPM results
are compared to the corresponding results obtained from an
unrestricted shell model calculation with the orbitals 
$1d_{5/2}0g_{7/2}1d_{3/2}2s_{1/2}0h_{11/2}$ defining the shell model
space, having in mind the discussion in section \ref{sec:sec3}. 

\subsection{Microscopic quasiparticle-phonon model}

In the extreme quasiparticle-phonon picture the two energies, ($E$(phen)
and $E$(calc) in Tables \ref{tab:odd-2}-\ref{tab:odd-7}, are obtained in 
the following way. 
For the phenomenological energy, $E$(phen), we have taken the phonon energy from experiments
(i.e. from the measured spectrum of the even-even $N=82$ isotone with 
one less proton) and the quasiparticle energy was deduced from the 
proton single-particle energies of Fig.\ \ref{fig:sp-energies} 
using the BCS expression
\begin{equation}
      E_{\alpha}^{\rm qp}=\sqrt{(\varepsilon_{\alpha}-\lambda_{\rm p})^2+
      \Delta^2},
      \label{eq:gap}
\end{equation}
where $\varepsilon_{\alpha}$ is the single-particle energy of the active
proton orbital and $\lambda_{\rm p}$ is the chemical potential for the 
protons. The chemical potential we have taken simply to correspond to the
energy of the last occupied proton orbital in the even-even $N=82$
reference nuclei. One exception is the case of $^{141}$Pr where in the
corresponding reference nucleus $^{140}$Ce the proton g$_{7/2}$ orbital
is completely filled and the chemical potential thus lies somewhere
between the g$_{7/2}$ and d$_{5/2}$ orbitals. In this case one can
deduce the value of the chemical potential by using the above formula
to reproduce the experimental energy difference between the $7/2^+$
and $5/2^+$ (single-quasiparticle) states in $^{141}$Pr.
The quantity $\Delta$ in Eq.\ (\ref{eq:gap}) is the pairing gap and assumes the
phenomenological value
\begin{equation}
     \Delta = 12A^{-1/2} {\rm MeV}.
\end{equation}
\begin{table}[htbp]
\begin{center}
\caption{Low-lying states for $^{135}$I. Experimental angular
momentum values in parentheses are tentative. Energies are given in MeV.
The symbol sqp in the leading configuration column denotes a
single-quasiparticle state.
For further information concerning the extreme qp-phonon picture, see the
text.}
\begin{tabular}{cc|cccc|c}
\hline
& & \multicolumn{3}{c}{Extreme qp-phonon picture} & & \\
$J^{\pi}$ & $E$(exp) & proposed config. & $E$(phen) & $E$(calc) 
& $E$(MQPM) & $E$(SM) \\ 
\hline
$7/2^{+}$ & 0.000 & $7/2^+$(sqp) & 0.00 & 0.00 & 0.000 & 0.000 \\
($5/2^{+}$) & 0.604 & $5/2^+$(sqp) & 0.38 & 0.70 & 0.649 & 0.508 \\
($5/2^{+}$) & 0.870 & $2^+_1\otimes 7/2^+$ & 1.28 & 1.51 & 1.405 & 0.888 \\
& & $2^+_1\otimes 7/2^+$ & 1.28 & 1.51 & 1.498($9/2^+$) & 1.349($9/2^+$) \\
& & $2^+_1\otimes 7/2^+$ & 1.28 & 1.51 & 1.533($3/2^+$) & 1.053($3/2^+$) \\
$11/2^{+}$ & 1.134 & $2^+_1\otimes 7/2^+$ & 1.28 & 1.51 & 1.543 & 1.322 \\
& & $2^+_1\otimes 7/2^+$ & 1.28 & 1.51 & 1.560($7/2^+$) & 1.619($7/2^+$) \\
$15/2^{+}$ & 1.422 & $4^+_1\otimes 7/2^+$ & 1.57 & 1.85 & 1.886 & 1.683 \\
$17/2^{+}$ & 1.994 & $6^+_1\otimes 7/2^+$ & 1.69 & 1.94 & 2.033 & 1.966 \\
\hline
\end{tabular}
\label{tab:odd-2}
\end{center}
\end{table}
\begin{table}[htbp]
\begin{center}
\caption{Low-lying states of $^{137}$Cs. Legend as in Table \ref{tab:odd-2}.}
\begin{tabular}{cc|cccc|c}
\hline
& & \multicolumn{3}{c}{Extreme qp-phonon picture} & & \\
$J^{\pi}$ & $E$(exp) & proposed config. & $E$(phen) & $E$(calc) 
& $E$(MQPM) & $E$(SM) \\ 
\hline
$7/2^{+}$ & 0.000 & $7/2^+$(sqp) & 0.00 & 0.00 & 0.000 & 0.000 \\
$5/2^{+}$ & 0.456 & $5/2^+$(sqp) & 0.38 & 0.45 & 0.404 & 0.208 \\
& & $2^+_1\otimes 7/2^+$ & 1.31 & 1.50 & 1.315($5/2^+$) & 0.994($5/2^+$) \\
& & $2^+_1\otimes 7/2^+$ & 1.31 & 1.50 & 1.471($9/2^+$) & 1.463($9/2^+$) \\
& & $2^+_1\otimes 7/2^+$ & 1.31 & 1.50 & 1.512($3/2^+$) & 1.233($3/2^+$) \\
& & $2^+_1\otimes 7/2^+$ & 1.31 & 1.50 & 1.529($7/2^+$) & 1.519($7/2^+$) \\
& & $2^+_1\otimes 7/2^+$ & 1.31 & 1.50 & 1.530($11/2^+$) & 1.422($11/2^+$) \\
$1/2^{+}$ & 1.490 & $2^+_1\otimes 5/2^+$,\ $4^+_1\otimes 7/2^+$ 
& 1.69 & 1.95-1.96 & 1.923 & 1.361 \\
$9/2^{-}$ & 1.868 & $2^+_1\otimes 11/2^-$,\ $3^-_1\otimes 7/2^+$ 
& 3.26-3.27 & 3.10-3.62 & 3.126 & 2.061 \\
$1/2^{+}$ & 2.150 & $1/2^+$(sqp) & 2.13 & 2.23 & 1.986 & 2.093 \\
\hline
\end{tabular}
\label{tab:odd-3}
\end{center}
\end{table}
\begin{table}[htbp]
\begin{center}
\caption{Low-lying states of $^{139}$La. Legend as in Table \ref{tab:odd-2}.
The underlined spin assignments are considered to be favoured by the SM
and the MQPM calculations.}
\begin{tabular}{cc|cccc|c}
\hline
& & \multicolumn{3}{c}{Extreme qp-phonon picture} & & \\
$J^{\pi}$ & $E$(exp) & proposed config. & $E$(phen) & $E$(calc) 
& $E$(MQPM) & $E$(SM) \\ 
\hline
$7/2^{+}$ & 0.000 & $7/2^+$(sqp) & 0.00 & 0.00 & 0.000 & 0.000 \\
$5/2^{+}$ & 0.166 & $5/2^+$(sqp) & 0.38 & 0.19 & 0.154 & 0.061 \\
$1/2^{+}$ & 1.209 & $2^+_1\otimes 5/2^+$\ ($4^+_1\otimes 7/2^+$) 
& 1.82\ (1.90) & 1.69\ (2.08) & 1.633 & 1.331 \\
$9/2^{+}$ & 1.219 & $2^+_1\otimes 7/2^+$ & 1.44 & 1.50 & 1.451 & 1.432 \\
$(5/2)^{+}$ & 1.257 & $2^+_1\otimes 7/2^+$ & 1.44 & 1.50 & 1.256 & 0.983 \\
($9/2^{+}$) & 1.381 & $2^+_1\otimes 5/2^+$ & 1.82 & 1.69 & 1.719 & 1.491 \\
$(11/2)^{-}$ & 1.420 & $11/2^-$(sqp) & 1.95 & 1.71 & 1.660 & 1.770 \\
$\underline{5/2^{+}},7/2^{+}$ & 1.421 & $2^+_1\otimes 5/2^+$ & 1.82 & 
1.69 & 1.644 & 1.467 \\
($9/2^{+}$) & 1.476 & $4^+_1\otimes 7/2^+$ & 1.90 & 2.08 & 2.068 & 1.818 \\
$11/2^{+}$ & 1.534 & $2^+_1\otimes 7/2^+$ & 1.44 & 1.50 & 1.530 & 1.389 \\
$7/2^{+}$ & 1.538 & $2^+_1\otimes 7/2^+$ & 1.44 & 1.50 & 1.462 & 1.530 \\
$\underline{3/2^{+}},5/2^{+}$ & 1.558 & $2^+_1\otimes 7/2^+$ & 1.44 & 
1.50 & 1.495 & 1.037 \\
$(15/2)^{-}$ & 3.375 & $2^+_1\otimes 11/2^-$ & 3.39 & 3.21 & 3.242 & 3.193 \\
$17/2^{-}$ & 3.927 & $4^+_1\otimes 11/2^-$ & 3.85 & 3.79 & 3.311 & 3.798 \\
\hline
\end{tabular}
\label{tab:odd-4}
\end{center}
\end{table}
\begin{table}[htbp]
\begin{center}
\caption{Low-lying states of $^{141}$Pr. Legend as in Table \ref{tab:odd-2}.}
\begin{tabular}{cc|cccc|c}
\hline
& & \multicolumn{3}{c}{Extreme qp-phonon picture} & & \\
$J^{\pi}$ & $E$(exp) & proposed config. & $E$(phen) & $E$(calc) 
& $E$(MQPM) & $E$(SM) \\ 
\hline
$5/2^{+}$ & 0.000 & $5/2^+$(sqp) & 0.00 & 0.00 & 0.000 & 0.000 \\
$7/2^{+}$ & 0.145 & $7/2^+$(sqp) & 0.145$^{\ 1)}$ & 0.072 & 0.104 & 0.008 \\
$11/2^{-}$ & 1.118 & $11/2^-$(sqp) & 1.31 & 1.31 & 1.301 & 1.497 \\
$3/2^{+}$ & 1.127 & $2^+_1\otimes 5/2^{+}$ & 1.60 & 1.48 & 1.314 & 0.936 \\
$(5/2)^{+}$ & 1.293 & $2^+_1\otimes 5/2^+$ & 1.60 & 1.48 & 1.371 & 0.993 \\
$1/2^{+}$ & 1.299 & $1/2^+$(sqp) & 1.49 & 1.42 & 1.396 & 1.148 \\
$3/2^{+}$ & 1.436 & $3/2^+$(sqp) & 1.23 & 1.30 & 1.342 & 1.140 \\
$(7/2)^{+}$ & 1.452 & $2^+_1\otimes 5/2^+$ & 1.60 & 1.48 & 1.440 & 1.452 \\
$9/2^{+}$ & 1.457 & $2^+_1\otimes 5/2^+$ & 1.60 & 1.48 & 1.522 & 1.413 \\
$11/2^{+}$ & 1.494 & $2^+_1\otimes 7/2^+$ & 1.74 & 1.55 & 1.622 & 1.418 \\
$9/2^{+}$ & 1.521 & $2^+_1\otimes 7/2^+$ & 1.74 & 1.55 & 1.549 & 1.413 \\
$(3/2)^{+}$ & 1.608 & $2^+_1\otimes 7/2^+$ & 1.74 & 1.55 & 1.626 & 1.372 \\
$1/2^{+}$ & 1.658 & $2^+_1\otimes 5/2^+$ & 1.60 & 1.48 & 1.602 &1.545  \\
$13/2^{+}$ & 1.768 & $4^+_1\otimes 5/2^+$ & 2.08 & 2.14 & 2.206 & 1.939 \\
$15/2^{+}$ & 1.797 & $6^+_1\otimes 5/2^+$ & 2.11 & 2.21 & 2.278 & 1.978 \\
$13/2^{+}$ & 1.986 & $6^+_1\otimes 5/2^+$ & 2.11 & 2.21 & 2.206 & 2.004 \\
$17/2^{+}$ & 2.070 & $6^+_1\otimes 5/2^+$ & 2.11 & 2.21 & 2.280 & 2.064 \\
\hline
\end{tabular}
1) The chemical potential of Eq.\ (19) was fixed by the experimental 
quasiparticle energy. 
\label{tab:odd-5}
\end{center}
\end{table}
\begin{table}[htbp]
\begin{center}
\caption{Low-lying states of $^{143}$Pm. Legend as in Table \ref{tab:odd-2}.
The underlined spin assignments are considered to be favoured by the SM
and the MQPM calculations.}
\begin{tabular}{cc|cccc|c}
\hline
& & \multicolumn{3}{c}{Extreme qp-phonon picture} & & \\
$J^{\pi}$ & $E$(exp) & proposed config. & $E$(phen) & $E$(calc) 
& $E$(MQPM) & $E$(SM) \\ 
\hline
$5/2^{+}$ & 0.000 & $5/2^+$(sqp) & 0.00 & 0.00 & 0.000 & 0.000 \\
$7/2^{+}$ & 0.272 & $7/2^+$(sqp) & 0.39 & 0.32 & 0.342 & 0.003 \\
$11/2^{-}$ & 0.960 & $11/2^-$(sqp) & 1.08 & 1.01 & 1.009 &1.203  \\
$3/2^{+}$ & 1.060 & $3/2^+$(sqp) & 1.01 & 1.06 & 1.088 & 0.728 \\
$1/2^{+}$ & 1.173 & $1/2^+$(sqp) & 1.26 & 1.14 & 1.166 & 0.731 \\
($\underline{3/2},5/2$) & 1.287 & $2^+_1\otimes 5/2^+$ & 1.58 & 
1.46 & 1.176 & 1.391 \\
$3/2^{+}$ & 1.403 & $2^+_1\otimes 7/2^+$ & 1.96 & 1.79 & 1.854 &  \\
$9/2^{+}$ & 1.456 & $2^+_1\otimes 5/2^+$ & 1.58 & 1.46 & 1.531 & 1.512 \\
$3/2^{+},\underline{5/2^{+}}$ & 1.515 & $2^+_1\otimes 5/2^+$ & 1.58 & 
1.46 & 1.524 & 1.325 \\
($5/2^{+}$) & 1.566 & $2^+_1\otimes 5/2^+$ & 1.58 & 1.46 & 
1.521($7/2^+$) & 1.430($7/2^+$) \\
$(9/2)^{+}$ & 1.566 & $2^+_1\otimes 7/2^+$  & 1.96 & 1.79 & 1.747 &  \\
$3/2^{+},\underline{5/2^{+}}$ & 1.614 & $2^+_1\otimes 7/2^+$ & 1.96 & 
1.79 & 1.680 & 1.403 \\
$11/2^{+}$ & 1.664 & $2^+_1\otimes 7/2^+$ & 1.96 & 1.79 & 1.854 &  \\
$1/2^{+}$ & 1.753 & $2^+_1\otimes 5/2^+$ & 1.58 & 1.46 & 1.537 &1.632   \\
$15/2^{+}$ & 1.898 & $4^+_1\otimes 7/2^+$ & 2.49 & 2.46 & 2.528 &  \\
$17/2^{+}$ & 2.288 & $6^+_1\otimes 5/2^+$ & 2.21 & 2.36 & 2.432 &  \\
\hline
\end{tabular}
\label{tab:odd-6}
\end{center}
\end{table}
\begin{table}[htbp]
\begin{center}
\caption{Low-lying states of $^{145}$Eu. Legend as in Table \ref{tab:odd-2}. 
In this
case only the MQPM results are available since the SM calculations are
already beyond any reasonable effort.}
\begin{tabular}{cc|ccc|c}
\hline
& & \multicolumn{3}{c}{Extreme qp-phonon picture} & \\
$J^{\pi}$ & $E$(exp) & proposed configuration(s) & $E$(phen) & $E$(calc) 
& $E$(MQPM) \\ 
\hline
$5/2^{+}$ & 0.000 & $5/2^+$(sqp) & 0.00 & 0.00 & 0.000 \\
$7/2^{+}$ & 0.329 & $7/2^+$(sqp) & 0.39 & 0.54 & 0.551 \\
$11/2^{-}$ & 0.716 & $11/2^-$(sqp) & 1.08 & 0.60 & 0.618 \\
$1/2^{+}$ & 0.809 & $1/2^+$(sqp) & 1.26 & 0.76 & 0.788 \\
$3/2^{+}$ & 1.042 & $3/2^+$(sqp) & 1.01 & 0.71 & 0.737 \\
$9/2^{-}$ & 1.368 & $3^-_1\otimes 5/2^+$ & 1.81 & 1.48 & 1.542 \\
$1/2^{+}$ & 1.460 & $2^+_1\otimes 5/2^+$ & 1.66 & 1.48 & 1.547 \\
$7/2^{-}$ & 1.500 & $3^-_1\otimes 5/2^+$ & 1.81 & 1.48 & 1.541 \\
& & $3^-_1\otimes 5/2^+$ & 1.81 & 1.48 & 1.532($1/2^-$) \\
$3/2^{-}$, $5/2^{-}$ & 1.567 & $3^-_1\otimes 5/2^+$ & 1.81 & 1.48 & 
1.542($5/2^-$) \\
$3/2^{-}$ & 1.600 & $3^-_1\otimes 5/2^+$ & 1.81 & 1.48 & 1.535 \\
$11/2^{-}$ & 1.602 & $3^-_1\otimes 5/2^+$ & 1.81 & 1.48 & 1.553 \\
& & $2^+_1\otimes 5/2^+$ & 1.66 & 1.48 & 1.543($9/2^+$) \\
& & $2^+_1\otimes 5/2^+$ & 1.66 & 1.48 & 1.571($7/2^+$) \\
& & $2^+_1\otimes 5/2^+$ & 1.66 & 1.48 & 1.587($5/2^+$) \\
$7/2^{-}$ & 1.745 & $3^-_1\otimes 7/2^+$, $2^+_1\otimes 11/2^-$ 
& 2.20,2.57 & 2.02,2.08 & 2.080 \\
$3/2^{+}$ & 1.758 & $2^+_1\otimes 5/2^+$ & 1.66 & 1.48 & 1.258 \\
$3/2^{-}$ & 1.762 & $3^-_1\otimes 7/2^+$ & 2.20 & 2.02 & 2.070 \\
$5/2^{-}$ & 1.766 & $3^-_1\otimes 7/2^+$ & 2.20 & 2.02 & 2.065 \\
$11/2^{-}$ & 1.792 & $3^-_1\otimes 7/2^+$, $2^+_1\otimes 11/2^-$ 
& 2.20,2.57 & 2.02,2.08 & 2.084 \\
$9/2^{-}$ & 1.827 & $3^-_1\otimes 7/2^+$, $2^+_1\otimes 11/2^-$ 
& 2.20,2.57 & 2.02,2.08 & 2.084 \\
& & $2^+_1\otimes 7/2^+$, $3^-_1\otimes 11/2^-$ 
& 2.05,2.89 & 2.02,2.07 & 1.928($7/2^+$) \\
$3/2^{+}$,$5/2^{+}$ & 1.845 & $2^+_1\otimes 7/2^+$, $3^-_1\otimes 11/2^-$ 
& 2.05,2.89 & 2.02,2.07 & 1.952($5/2^+$) \\
& & $2^+_1\otimes 7/2^+$, $3^-_1\otimes 11/2^-$ 
& 2.05,2.89 & 2.02,2.07 & 1.977($9/2^+$) \\
$1/2^{+}$, $3/2^{+}$ & 1.881 & $2^+_2\otimes 5/2^+$ & 2.42 & 2.58 & 
2.090($1/2^+$) \\
$(3/2,5/2)^{+}$ & 1.915 & $2^+_1\otimes 7/2^+$, $3^-_1\otimes 11/2^-$ 
& 2.05,2.89 & 2.02,2.07 & 2.161($5/2^+$) \\
$3/2^{+}$ & 2.049 & $2^+_1\otimes 7/2^+$ & 2.05 & 2.02 & 2.295 \\
$5/2^{+}$ & 2.114 & $2^+_1\otimes 7/2^+$ & 2.05 & 2.02 & 2.252 \\
$9/2^{-}$,$11/2^{-}$ & 2.117 & $3^-_1\otimes 7/2^+$, $2^+_1\otimes 11/2^-$ 
& 2.20,2.57 & 2.02,2.08 & 2.095($9/2^-$) \\
\hline
\end{tabular}
\label{tab:odd-7}
\end{center}
\end{table}
The calculated extreme quasiparticle-phonon energy, $E$(calc), 
was obtained
by summing the QRPA-calculated energy (see Ref.\ \cite{HOL97}) 
of the $J^{\pi}$ phonons of the even reference nuclei 
with the single-quasiparticle
energy coming from the BCS calculation.

In the following we shall comment on the MQPM results 
shown in Tables \ref{tab:odd-2}-\ref{tab:odd-7}.
The emphasis lies on analyzing
the lowest-lying quasiparticle-phonon multiplets of the odd isotones
with respect to their span in energy, their 
centroids and the relative
location of the various spins within the perturbed multiplet. 
The multiplet of the extreme quasiparticle-phonon 
picture looses its energy degeneracy through the 
action of the residual interactions and 
the resulting multiplet we call a perturbed one. It is of interest
to see how the breaking of the degeneracy evolves 
from nucleus to nucleus
and what its characteristic features are.

We start first by noticing that the residual 
interactions yield relatively
little perturbation on the energies of the 
single-quasiparticle-type states 
(denoted by the symbol (sqp) in the tables) in the MQPM calculation. 
The agreement in the single-quasiparticle energies
between the MQPM and experiment is good 
due to the success of the  BCS 
calculation in describing the lowest excitations of the odd isotones.
This, in turn, means that the 
phenomenological mean field, represented by
the proton single-particle energies of Fig.\ \ref{fig:sp-energies}, 
is consistent with the
residual pairing interactions used in the present calculation.

As a general feature of the MQPM results one can say that the MQPM
describes better the heavier $N=82$ 
isotones (both even \cite{HOL97} and
odd) since the heavier isotones are not in the immediate vicinity of
the $Z=50$ closed core and thus the quasiparticle description is 
expected to be better justified.

For $^{135}$I the wrong theoretical $2_1^+$ 
energy in $^{134}$Te (see
Ref.\ \cite{HOL97}) leads to too high an energy for the multiplet
$2^+_1\otimes 7/2^+$ (see Table \ref{tab:odd-2}) 
and thus to too high an energy for
the associated MQPM states (i.e. to too high a centroid of the MQPM 
multiplet). The same is valid for the $4^+_1$ energy and the associated
$4^+_1\otimes 7/2^+$ multiplet. For the
other even reference isotones the phonon energies of the $2^+_1,3^-_1,
4^+_1$ and $6^+_1$ states are in reasonable agreement with experiment 
and in the case of $^{138}$Ba, $^{140}$Ce and $^{142}$Nd the agreement
is rather good \cite{HOL97}.

Let us now discuss the lowest observed multiplets of the odd isotones,
namely $2^+_1\otimes 7/2^+$ for $A=139$ and $2^+_1\otimes 5/2^+$ 
for $A=141,143$, as well as $3^-_1\otimes 5/2^+$ for $^{145}$Eu. 
Since experimental data on the $2^+_1\otimes 7/2^+$ multiplet
is missing or incomplete for $^{135}$I and $^{137}$Cs,
only $^{139}$La is left for comparison. From
Table \ref{tab:odd-4} one observes that the experimental and the MQPM 
centroids correspond
to each other rather nicely and the width of both spectra is the same.
In addition, the $3/2^+$, $5/2^+$, $7/2^+$ and $11/2^+$ members of the
multiplet are reproduced by the MQPM rather well but the MQPM clearly
fails for the $9/2^+$ state. In this case the phenomenological
quasiparticle-phonon energy $E({\rm phen})=1.44$MeV is more or less
the centroid of both the experimental and the MQPM multiplet. The
corresponding calculated energy $E({\rm calc})=1.50$MeV is slightly
higher.

The above described features are also seen in the $2^+_1\otimes 7/2^+$ 
multiplet in $^{141}$Pr, where this multiplet is not any more the lowest
one but above the $2^+_1\otimes 5/2^+$ multiplet 
(see Table \ref{tab:odd-5}). A clear
agreement of the MQPM with the data, especially
in the case of the $3/2^+$ member of the multiplet, is evident.

The same type of analysis can be performed for the $2^+_1\otimes 5/2^+$ 
multiplet which is the lowest one in $^{141}$Pr and $^{143}$Pm (see Tables
\ref{tab:odd-5} and \ref{tab:odd-6} ). 
In Table \ref{tab:odd-5} it is seen that for
$^{141}$Pr the centroid of the experimental multiplet is well reproduced
by the MQPM as is also the case for $^{143}$Pm. A closer look at
the spectrum of $^{141}$Pr reveals that the $3/2^+$
and $9/2^+$ states are too far away from the $7/2^+$ state and the
states are more homogeneously distributed in the theoretical spectrum.
The undisturbed quasiparticle-phonon energies, $E$(phen) and $E$(calc),
lie near the top of the experimental and theoretical multiplet, 
respectively. This situation is reversed in the case of $^{143}$Pm, namely
now the experimental multiplet is more homogeneously distributed due
to the fact that the MQPM fails in predicting the location of the
$9/2^+$ state whereas for the other members of the multiplet the MQPM
energies roughly correspond to the experimental ones even though the
$5/2^+$ and $7/2^+$ states are inverted in the theory. Once again the
undisturbed quasiparticle-phonon energies are near the top of the perturbed
multiplet.

Finally, in Table \ref{tab:odd-7} 
one can observe the lowest multiplet,
$3^-_1\otimes 5/2^+$, of $^{145}$Eu. One can see that the centroid of the
calculated multiplet is more or less correct but that the multiplet
is far too compressed when compared with the experimental span of the
multiplet (in experiment the $9/2^-$ state comes very much down in
energy). It seems that in the MQPM the biggest qualitative and 
quantitative problems appear for the $3/2^-$ and $9/2^-$ members of
the multiplet. It has to be noted that in Table \ref{tab:odd-7} 
the labelling
of the higher excited states by the extreme quasiparticle-phonon wave function
is to be taken with a grain of salt since many of the MQPM states
at these energies can be constructed by superposition of a large number
of possible quasiparticle-phonon components lying rougly at the same
energy region. Thus the overlap of the proposed quasiparticle-phonon
structures of Table \ref{tab:odd-7} with the MQPM wavefunctions is not
necessarily very big. The assignments may be considered more as
a means of keeping track of the experimental states with the same spin.

As it was stated earlier, the spreading of the multiplet of the
extreme quasipar\-ti\-cle-phonon picture can be ascribed to the influence
of the residual interactions, coming mainly from the $H_{31}$ part
of the residual hamiltonian. This spreading in energy ranges roughly
from 100 keV to 400 keV in the discussed odd isotones and the 
calculated extreme quasiparticle-phonon energy ($E$(calc) in the tables)
lies rather close to the top of the perturbed multiplet. This means that
the residual interactions tend to redistribute the multiplet energies
towards lower energies (the centroid of the perturbed multiplet is always
substantially lower than the extreme quasiparticle-phonon energy). 

It is to be noted also that in the MQPM calculation the energies of 
the $3/2^+$ and $9/2^+$ members of the $2^+_1\otimes 7/2^+$ multiplet
are the ones least affected by the residual interactions in the case of
$^{135}$I, $^{137}$Cs, $^{139}$La and $^{141}$Pr.
For the $2^+_1\otimes 5/2^+$ multiplet
all the spin members  of the multiplet are clearly affected by the residual
interaction, some of them considerably. The most strongly 
affected member of the
$2^+_1\otimes 7/2^+$ multiplet is the $5/2^+$ state whereas $3/2^+$ state
is most affected in the $2^+_1\otimes 5/2^+$ multiplet.

\subsection{Comparison with shell model results}

In the MQPM scheme one starts with the BCS occupation 
probabilities stemming from the even isotones, given
in Ref.\ \cite{HOL97}. As discussed in section \ref{sec:sec2}, the
residual many-body hamiltonian can be expressed in terms of these
occupation probabilities. In general, for the single-particle states
we obtained a good agreement in Ref.\ \cite{HOL97} 
between the shell model occupation
probabilities and those of a BCS calculation 
for the even isotones. Since the states of the MQPM, discussed
in Tables \ref{tab:odd-2}-\ref{tab:odd-7}, are combinations
of single-particle states and phonons from the 
corresponding even isotones, it is then of interest
to see whether one can retrace eventual discrepancies 
between the shell model approach and the MQPM to the fact
that the SM employs a smaller set of single-particle
orbitals in the diagonalization than the MQPM.
It is therefore important to see how a renormalized
effective interaction obtained by perturbative many-body
methods is able to account for degrees of freedom 
not accounted for by the SM model space.  
It is, however, important to notice that although the
MQPM employs a larger set of single-particle orbitals, only 
a limited set of states are obtained from the 
diagonalizations. In the SM all states are, in principle,
taken into account in the diagonalization.
  
The shell model results are presented in Tables 
\ref{tab:odd-2}-\ref{tab:odd-6}.
In, general, there is a fairly good agreement 
between the shell model results and experiment, with
deviations of the order of $100-300$ keV. However, as the
number of valence particles increases from $3$ in 
$^{135}$I to $13$ in $^{143}$Pm, the description of the 
spacing between the lowest-lying $5/2^+$ and $7/2^+$ states gets worse. 

In order to understand these differences we have performed 
additional shell model calculations for $^{141}$Pr
with just the $G$-matrix as effective interaction,
in order to see whether different approaches
to the effective interaction within the 
$2s1d0g_{7/2}0h_{11/2}$ model space yield 
significant discrepancies. These results are displayed
in Table \ref{tab:shellmqpm}, under the column labelled 
$E(G)$. The shell model results with the effective
interaction to third-order in $G$ from Table \ref{tab:odd-5},
together with the corresponding MQPM results, are
included for comparison. These are labelled by $E$(SM) and
$E$(MQPM), respectively. 

One can note from Table \ref{tab:shellmqpm} that when going from the
third-order effective interaction over to the bare G-matrix interaction
in the shell model, the
spacing between the lowest-lying $5/2^+$ and $7/2^+$  states
increases to $0.034$ MeV, although it is still far from
the experimental value. The spectrum of the other states is more compressed
than for the $E$(SM) results. The proton
occupation numbers do also change, but all states of $^{141}$Pr are still
strongly dominated by  admixtures from the 
$5/2^+$ and $7/2^+$ single-particle states. 

\begin{table}[htbp]
\begin{center}
\caption{Lowest-lying states of given multipolarity in $^{141}$Pr 
using various approximations to the effective interaction.}
\begin{tabular}{ccccccc}
$J^{\pi}$ & $E$(exp) & $E$(BCS-1) & $E$(BCS-2) 
& $E$(MQPM) & $E$($G$) & $E$(SM) \\ 
\hline
$5/2^{+}$ & 0.000 & 0.17 & 0.00  & 0.00  & 0.00  & 0.000 \\
$7/2^{+}$ & 0.145 & 0.00 & 0.07  & 0.104 & 0.034 & 0.008 \\
$11/2^{-}$& 1.118 & 1.42 & 1.31  & 1.301 & 0.889 & 1.497 \\
$3/2^{+}$ & 1.127 & 1.44 & 1.30  & 1.314 & 0.915 & 0.936 \\
$1/2^{+}$ & 1.299 & 1.58 & 1.42  & 1.396 & 1.292 & 1.148 \\
\hline
\end{tabular}
\label{tab:shellmqpm}
\end{center}
\end{table}

The question then arises whether these differences can be traced back to
the use of a smaller model space in the shell model calculation.
We have therefore performed a BCS calculation
in the model-space defined by the 
$2s1d0g_{7/2}0h_{11/2}$ single-particle orbitals, using the 
$G$-matrix as interaction and the 
single-particle energies of Fig. \ref{fig:sp-energies}. These results 
are labelled $E$(BCS-1) in Table \ref{tab:shellmqpm}. It has to be
noted that in this BCS calculation the mean-field part is the same 
as in the $E(G)$ calculation but that
a  BCS calculation includes only a very restricted set of states
as compared to the full shell model diagonalization.
The comparison reveals then how good or bad the BCS approximation is.
As stated earlier, the BCS calculation forms the basis for the
MQPM method. Moreover, the MQPM method used here employs 
a larger single-particle basis than that used in the shell model calculation
or the BCS-1 calculation. It may therefore be of interest to see
how the BCS calculation changes when we go to the model space
employed in the MQPM calculation. The results of such a BCS calculation
are denoted by  $E$(BCS-2) in Table \ref{tab:shellmqpm}.

One can see from Table \ref{tab:shellmqpm} that the results of 
the BCS-1 calculation clearly
deviate from the BCS-2 results, indicating the importance
of a larger single-particle basis. The relative positions of the 
$5/2^+$ and $7/2^+$ states are inverted and the other states 
are higher up in excitation energy in the BCS-1 calculation. 
Comparing the BCS-1 results with the 
$E(G)$ results, which are actually in reasonable
agreement with experiment, clearly indicates that the
simple BCS picture is far from sufficient in the restricted
$2s1d0g_{7/2}0h_{11/2}$ valence space.
The BCS-2 calculation, which employs the larger set of single-particle
states, yields results which are closer to experiment and
close to the MQPM results. The mere difference between the BCS-2
and MQPM results is in the better 
reproduction of the $5/2^+-7/2^+$ spacing by the MQPM.

In summary, the results in Table \ref{tab:shellmqpm} seem to indicate that
degrees of freedom not accounted for by the model space employed
in the shell model calculation, are important in order to get 
a proper reproduction of the experimental  spacing between
the two lowest-lying $5/2^+$ and $7/2^+$ states. 
However, when comparing the BCS-1 results with those of a shell model
calculation with the bare $G$-matrix, one sees that there are 
important differences. Sources of these discrepancies
are the many-body configurations not accounted for by the
BCS approach. How these differences will appear in a shell model
calculation which would employ the same set of single-particle
energies as the BCS-2 or MQPM approaches, is however not clear. 
  
In column three of Tables \ref{tab:odd-2} -- \ref{tab:odd-7} we indicate
the MQPM states that are supposed to be of single-quasiparticle nature.
For comparison with the shell model we create single--quasiparticle states
for the odd-nucleon system by coupling a $(j^{\pi})$ particle to the 
$0^{+}$ ground state of the neighbouring $A-1$ even system, 
$a^{\dagger}_{j}|{\rm SM}(A-1)\rangle $. 
By calculating the squared overlap between
the constructed single--quasiparticle state and the SM state,
$\vert\langle j^{\pi};{\rm SM}(A)|a^{\dagger}_{j}|0^{+};{\rm SM}(A-1) 
\rangle \vert^{2}$, we obtain
a measure of the fraction of single-quasiparticle structure in our SM
wave function. The results are tabulated in Table \ref{tab:quasi-particle}.
\begin{table}[htbp]
\begin{center}
\caption{Squared overlaps 
$\vert\langle j^{\pi};{\rm SM}(A)|a^{\dagger}_{j}|0^{+};{\rm SM}(A-1) 
\rangle \vert^{2}$ for the SM and one-quasiparticle probabilities for the
MQPM.}
\begin{tabular}{lcccccccccc}
\hline
$J^{\pi}$ & \multicolumn{2}{c}{$^{135}$I} & \multicolumn{2}{c}{$^{137}$Cs}
& \multicolumn{2}{c}{$^{139}$La} & \multicolumn{2}{c}{$^{141}$Pr} 
& \multicolumn{2}{c}{$^{143}$Pm} \\
& SM & MQPM & SM & MQPM & SM & MQPM & SM & MQPM & SM & MQPM \\
\hline
$7/2^{+}_{1}$  & 0.99 & 0.99 & 0.99 & 0.99 & 0.98 & 0.99 & 0.97 & 0.99
& 0.95 & 0.99 \\
$5/2^{+}_{1}$  & 0.94 & 0.98 & 0.95 & 0.98 & 0.95 & 0.99 & 0.94 & 0.98
& 0.98 & 0.98 \\
$3/2^{+}_{1}$  & 0.00 & 0.01 & 0.01 & 0.03 & 0.07 & 0.19 & 0.12 & 0.45
& 0.89 & 0.97 \\
$3/2^{+}_{2}$  & 0.00 & 0.00 & 0.09 & 0.00 & 0.02 & 0.00 & 0.75 & 0.52
& 0.00 & 0.01 \\
$1/2^{+}_{1}$  & 0.11 & 0.00 & 0.11 & 0.15 & 0.29 & 0.32 & 0.89 & 0.73
& 0.90 & 0.97 \\
$11/2^{-}_{1}$ & 0.98 & 0.94 & 0.96 & 0.95 & 0.94 & 0.95 & 0.91 & 0.96
& 0.89 & 0.96 \\
\hline
\end{tabular}
\label{tab:quasi-particle}
\end{center}
\end{table}
There is a nice correspondence
between the MQPM states proposed to be of single--quasiparticle nature and 
those SM states with predominantly single--quasiparticle structure.
In the case of the $3/2^+_1$ and $3/2^+_2$ states in $^{141}$Pr there
is a strong mixing of the one- and three-quasiparticle components in
the MQPM, stronger than in the SM.
Squared overlaps of magnitudes 0.95 -- 1.00 confirm that the $5/2^{+}_{1}$
and the $7/2^{+}_{1}$ states are fairly pure single--quasiparticle states.

We then turn to a comparison of the predominantly three-quasiparticle 
states in the two models. For this we consider the multiplets
$2^+_1\otimes 7/2^+$ for $A=135,137,139$ and $2^+_1\otimes 5/2^+$ 
for $A=141,143$. We start with the $2^+_1\otimes 7/2^+$ 
multiplet. Looking at Tables \ref{tab:odd-2}-\ref{tab:odd-4} one can 
see differences in the SM and the MQPM spectra of states ($3/2^+$,
$5/2^+$, $7/2^+$, $9/2^+$, $11/2^+$) belonging to this multiplet.
Characteristic features are 1) the centroid of the SM
multiplet is always lower than the centroid of the MQPM multiplet.
2) The high-spin members of the multiplet ($7/2^+$, $9/2^+$, $11/2^+$) 
correspond quite well to each other in the SM and the MQPM whereas the
$5/2^+$ member of the multiplet is consistently lower in the SM spectra
than in the MQPM ones (in both cases the $5/2^+$ state is usually the
lowest one). 3) Differences show up for the lowest-spin member of the 
multiplet, i.e. for spin $3/2^+$, which is always the second lowest in 
energy in the SM spectra but among the three highest levels in the MQPM 
spectra.

The above listed properties of the SM and the MQPM spectra distinguish 
between the two calculations. For $^{135}$I and $^{137}$Cs the data is missing
or incomplete so that essentially only $^{139}$La is left for comparison. From
Table \ref{tab:odd-4} one observes that the experimental and the MQPM centroids
correspond to each other rather nicely and the width of both spectra is the 
same. In addition, the $3/2^+$, $5/2^+$, $7/2^+$ and $11/2^+$ members of the
multiplet are reproduced by the MQPM rather well (both the SM and the 
MQPM fail for the $9/2^+$ state) whereas in the case of the SM large 
deviations are observed for the $5/2^+$ member and especially for the
$3/2^+$ member of the multiplet. Thus the experiment favours the
MQPM sequence of levels in the $2^+_1\otimes 7/2^+$ multiplet (although
also in the MQPM the sequence of levels is not completely correct).
For the $2^+_1\otimes 7/2^+$ multiplet in $^{141}$Pr (see Table 
\ref{tab:odd-5}) the differences between the two models and the experimental 
data are small.

The same type of analysis can be performed for the $2^+_1\otimes 5/2^+$ 
multiplet which is the lowest one in $^{141}$Pr and $^{143}$Pm (see Tables
\ref{tab:odd-5} and \ref{tab:odd-6}). In this case the differences between 
the SM and the MQPM spectra are less than for the $2^+_1\otimes 7/2^+$ case. 
For $^{141}$Pr the centroid of the experimental multiplet is well reproduced
by the MQPM, clearly better than by the SM, whereas for $^{143}$Pm both
the SM and the MQPM have roughly the correct centroid. The largest
difference between the SM and the MQPM spectra is found in the location of
the $5/2^+$ member of the multiplet and the experimental data favours the
MQPM for the $5/2^+$ energy.


\section{Conclusions}\label{sec:sec5}

The present work discusses the theoretical interpretation of
low--energy excitations of odd $N=82$ isotones
between the mass numbers $A=135$ and $A=143$. The energy spectra
of these isotones have been calculated by using the microscopic
quasiparticle--phonon model (MQPM) and the results have been compared
with the extreme quasiparticle--phonon picture and the results of
a large-basis shell model calculation with $3-11$ valence protons
outside the doubly-magic $^{132}$Sn core. This work is a direct
continuation of our earlier work on even $N=82$ isotones \cite{HOL97} and the
same realistic, microscopic two--body $G$--matrix interaction
has been used in the present MQPM calculation as was used in the QRPA
calculation of the even isotones (in fact, the QRPA calculation is a
necessary prerequisite of the present MQPM calculation). 
Also in the shell model
calculation we use the same effective two-body matrix elements, 
derived from the above-mentioned $G$--matrix elements through many-body 
perturbation techniques, which were used for the even isotones.

Overall, the spectra of the odd isotones are described well by the MQPM
considering that no fitting of the interaction was done.
This feature may be traced back to the capability of 
the BCS approach in describing
excitations of one-quasiparticle type. From this one can conclude that
the pairing-type of residual interactions used in the MQPM are consistent
with the mean field extracted from the experimental single-particle
energies in $^{133}$Sb. The low-lying experimental levels can be
labeled by their assumed leading quasiparticle-phonon contributions
and their energies can be compared with the energies of the unperturbed
quasiparticle-phonon multiplets and the perturbed ones emerging from
the MQPM calculations. In the MQPM both the widths and the centroids 
of the perturbed multiplets, as well as the sequences of different spins 
within the multiplets, are described rather nicely. The shell model
describes most individual states very well, in many cases better than
the MQPM, but has difficulties in describing the centroids of the 
quasiparticle-phonon multiplets and energies of some members of these 
multiplets, particularly the $3/2^+$ states in the $2^+_1\otimes 7/2^+$ 
multiplets and the $5/2^+$ states in the $2^+_1\otimes 5/2^+$ multiplets.

The aim of the present work was to see how well a truly microscopic
model, based on quasiparticle-phonon coupling and realistic microscopic
$G$-matrix interactions, can describe the level
systematics of a set of heavy semi-magic nuclei. At the same time
the results of these calculations can be compared with results coming
from a large-scale shell model calculation with truly microscopic
effective interaction based on the same $G$-matrix which is used in the 
quasiparticle-phonon calculation. Considering that in both calculations
only very few parameters enter the calculation, the success of both models
is surprisingly good.\newline
\newline
\newline
This work has been supported by the NorFA (Nordic Academy for Advanced Study).
Support from the Research Council of Norway (Programme for 
Supercomputing)  is also acknowledged.

\end{document}